\documentclass[aps,prl,twocolumn,showpacs,floatfix]{revtex4}

\usepackage{bm,bbm,amsmath, amssymb,amsbsy, amsthm,graphicx,subfigure,color,txfonts,multirow}
\usepackage[framemethod=tikz]{mdframed}
\usepackage[colorlinks]{hyperref}
\hypersetup{
	bookmarksnumbered,
	pdfstartview={FitH},
	citecolor={blue},
	linkcolor={red},
	urlcolor={black},
	pdfpagemode={UseOutlines}
}

 \newcommand{\tr}[1]{\text{Tr}}
\newcommand{\ket}[1]{|#1\rangle}
\newcommand{\bra}[1]{\langle#1|}

\newcommand{\med}[1]{\langle{#1}\rangle}

\newtheorem{Rem}{Remark}
\newtheorem{Res}{Result}

\begin{document}

\title{Observable measure of quantum coherence in finite dimensional systems}

\author{Davide Girolami$^{1,2}$}
\email{davegirolami@gmail.com}
\affiliation{$^1\hbox{Clarendon Laboratory, Department of Physics, University of Oxford, Parks Road, Oxford OX1 3PU, United Kingdom}$ \\
$^2\hbox{Department of Electrical and Computer Engineering, National University of Singapore, 4 Engineering Drive 3, Singapore 117583}$ }

\begin{abstract}
Quantum coherence is the key resource for quantum technology, with applications in quantum optics, information processing, metrology and cryptography.  
Yet, there is no universally efficient method for quantifying coherence either in theoretical or in experimental practice. 
 I introduce a framework for measuring quantum coherence in finite dimensional systems. I define a theoretical measure 
which satisfies the reliability criteria established in the context of quantum resource theories. 
Then, I present an experimental scheme implementable with current technology which evaluates the quantum coherence of an unknown state of a $d$-dimensional system by performing two programmable measurements on an ancillary qubit, in place of the $O(d^2)$ direct measurements required by  full state reconstruction.  The result yields a benchmark for monitoring quantum effects in complex systems, e.g.  certifying non-classicality in quantum protocols 
 and probing the quantum behaviour of biological complexes.
\end{abstract}

\date{\today}

 
  \maketitle
{\it Introduction --} While harnessing quantum coherence is matter of routine in delivering quantum technology  \cite{nielsen,metrorev,ekertcry,simrev,biorev}, and the quantum optics rationale rests on creation and manipulation of coherence  \cite{walls}, there is no universally efficient route to  measure the amount of quantum coherence carried by the state of a system in dimension $d>2$.  It is customary to employ quantifiers tailored to the scenario of interest, i.e. of not general employability, expressed in terms of {\it ad hoc} entropic functions, correlators, or  functions of the off-diagonal density matrix coefficients (if available) \cite{plenio,noriwit,agata}.  \\
Quantum information theory provides the framework to address the problem. Physical laws are interpreted as restrictions on the accessible quantum states and operations, while the properties of physical systems are the resources that one must consume to perform a task under such laws \cite{opp}. An algorithmic characterization of quantum coherence as a resource and a set of {\it bona fide} criteria for coherence monotones have been identified \cite{plenio,aberg,luocri}. Also, coherence has been shown to be related to the asymmetry of a quantum state \cite{reviewW,newmar}.
 On the experimental side, the scalability of the detection scheme is a major criterion  in developing witnesses and measures of coherence, as we are interested in exploring the quantum features of highly complex macrosystems, e.g. multipartite quantum registers and networks. Therefore, it is desirable to have a coherence measure which is both theoretically sound and experimentally appealing.\\

Here I introduce a measure of quantum coherence for states of finite dimensional systems. The quantity satisfies the properties of reliable coherence quantifiers and it is easy to compute, not involving any optimization. Also, it has a lower bound which is experimentally observable. The detection of quantum coherence does not require to reconstruct the full density matrix of the state, but it relies upon the estimation of quadratic functionals of the density matrix coefficients. I propose a scheme 
which is readily implementable with current quantum technology, e.g. in all-optical setup \cite{walls}.  Regardless of the dimensionality $d$ of the system, the protocol requires to realize two programmable measurements \cite{nielsen2,dariano,brun,paz,ekertdir,filip}, which are basic operations in quantum information, on an ancillary qubit which undergoes a unitary interaction with the system under scrutiny. An alternative scheme requiring $O(d)$ measurements overcomes the implementation of multipartite controlled gates.\\

\begin{figure} 
  \begin{mdframed}[leftmargin=0.3cm,rightmargin=0.3cm,
innerleftmargin=.5cm,innerrightmargin=.3cm,innertopmargin=0.5cm,innerbottommargin=.3cm]
 \includegraphics[width=\textwidth]{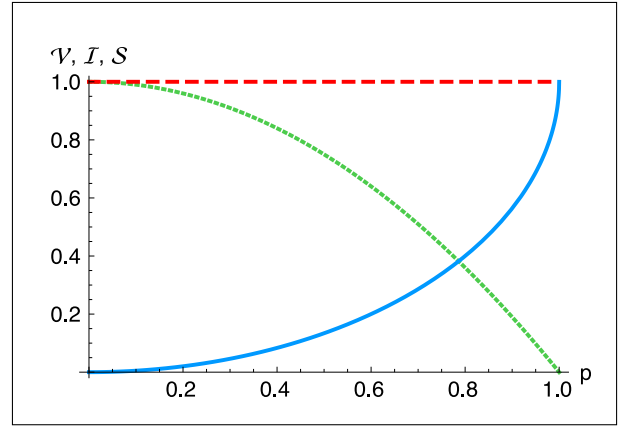}
 \end{mdframed}
 \caption[Coherence as quantum uncertainty]{(Colors online) Coherence as quantum uncertainty. A measurement implies two kinds of uncertainty. A classical one, which is quantified by the state mixedness and it is independent of the measured observable; a quantum contribution to the  uncertainty, which is observable-dependent and it reflects the quantum coherence of the state. The plot shows the uncertainty on the measurement of the observable  $\sigma_z =  \left( \begin{array}{cc}1 & 0  \\0 & -1  \end{array} \right)$ in the qubit $\rho =(1-p) \mathbb{I}_2/2 + p |\psi\rangle\bra{\psi}, \ket{\psi}=\frac{1}{\sqrt2}(\ket 0+\ket 1), p\in [0,1]$. The red dashed line is the variance of the $\sigma_z$ operator: ${\cal V}(\rho,\sigma_z)=\med{\sigma_z^{\ 2}}_\rho-\med{\sigma_z}_\rho^2$.  The blue continuous curve represents the quantum coherence ${\cal I}(\rho, \sigma_z)$. The green dotted curve depicts the linear entropy ${\cal S}(\rho)=2-2 \text{Tr}[\rho^2]$, which measures the classical uncertainty. As expected by a coherence measure, the skew information monotonically increases  with $p$.} 
  \label{cohex}
 \end{figure}

{\it Measure of coherence: theory --}  In a quantum measurement,  we observe wave-like probability distributions of outcomes. In particular, the uncertainty of a measurement is twofold \cite{luo,kamil}. First, an inherently classical indeterminacy is brought about by the ignorance  about the state of the system, being quantified by its mixedness. Second, a quantum uncertainty is due to the fact that the state is changed by the measurement. The quantum coherence of the state embodies the latter contribution to the unpredictability of the outcome. A state $\rho$ is left invariant by measuring an observable $K$ (assumed bounded and non-degenerate) if and only if it does not show coherence in the $K$ eigenbasis, being an eigenstate or a mixture of eigenstates of the observable, i.e. $[\rho,K]=0$ \cite{tufa}.   \\
A quantitative characterization to the above argument is the following.  I define the $K$-coherence of a $d$-dimensional state $\rho$ as the quantum coherence it carries when measuring $K$. For a pure state $\phi$, the uncertainty on the outcome, which is exclusively due to quantum coherence, can be safely measured by the variance ${\cal V}(|\phi\rangle,K)$. Given the spectral decomposition $K=k_i |k_i\rangle\langle k_i|$, we have ${\cal V}(|\phi\rangle,K)=\sum_i k_i^2(K_{i \phi}-K_{i \phi}^2) -\sum_{i\neq j} k_i k_j K_{i \phi}K_{j \phi}$,  which is a non-negative function of the coherence terms $K_{i \phi}=|\langle\phi |k_i\rangle|^2$.
For a mixed state $\rho=\sum_i p_i |\phi_i\rangle\langle \phi_i|, \sum_i p_i =1$, the situation is more complex.  The variance is now affected by the state mixedness. We can formally split it in a quantum and a classical part: ${\cal V}(\rho,K)={\cal V}^Q(\rho,K)+{\cal V}^C(\rho,K)$ \cite{luo}. Coherence is then related to the truly quantum share ${\cal V}^Q(\rho,K)$, obtained filtering out the uncertainty ${\cal V}^C(\rho,K)$ due to mixing. We search for a measure which is non-negative (it is a measure of uncertainty), zero if and only if states and observable commute (faithful), convex  (non-increasing under mixing), and bounding from below the variance, being equal to it for pure states.  A class of functions which enjoy all these properties is given by the Wigner-Yanase-Dyson skew informations \cite{wy}
\begin{eqnarray}
 {\cal V}^Q(\rho,K)={\cal I}^p(\rho,K)=-1/2 \text{Tr}\Big[ [\rho^p,K][\rho^{1-p},K] \Big], 0<p<1.
\end{eqnarray}
For technical convenience, I fix $p=1/2$ (from now on, the index is dropped) and prove that 
\begin{Res}\label{cohres}
The skew information ${\cal I}(\rho, K) =-1/2$\ \normalfont{Tr}$\Big[[\sqrt{\rho},K]^2\Big]$ is a measure of the $K$-coherence of the state $\rho$.
 \end{Res}
  Indeed, the skew information satisfies the {\it bona fide} criteria for coherence monotones  \cite{plenio,aberg,luocri} (see proof at the end).   It was originally introduced to quantify the quantum uncertainty in measurements under conservation laws \cite{wy}, and later investigated in quantum statistics \cite{hansen,isola,luo,luo1,brody} and characterization of quantum correlations \cite{luo2,tufa}. For mixed states, the skew information can be interpreted as the lower bound of the weighted statistical uncertainty about $K$ for any possible state preparation, i.e. ${\cal I}(\rho,K)\leq \sum_i p_i {\cal V}(|\phi_i\rangle,K), \forall \{\phi_i\}$.  
 A numerical example is presented in Fig.~\ref{cohex}. Consistently, given a $n$-partite system $A_{1,2,\ldots n}$, the local $K_{A_i}$-coherence is given by ${\cal I}(\rho_{A_1,A_2,\ldots, A_n},\mathbb{I}_{A_i,A_2,\ldots, A_{i-1}}\otimes K_{A_i}\otimes \mathbb{I}_{A_{i+1},A_{i+2},\ldots,A_n})$ \cite{notacorr}.\\
 It is noticeable that the skew informations yields a common framework for two quantum resources, i.e coherence and asymmetry.  The latter is the ability of a state to act as a reference frame under a superselection rule, being widely investigated in recent years \cite{reviewW,newmar,peres,acin,bart,wise,kitaev,vacca,gour,monotone,vacca2,gour2,cpt,exp,sciarrino,rudo}. One can observe that asymmetry is the quantum coherence lost by applying a phase shift w.r.t. the eigenbasis of a ``supercharge'' $Q$ \cite{gour,newmar}. Then, the quantity ${\cal I}(\rho, Q)$ turns out to be a full-fledged measure of asymmetry \cite{epaps}.   \\
 
\begin{figure} 
\begin{mdframed}[leftmargin=0.2cm,rightmargin=0.2cm,
innerleftmargin=1cm,innerrightmargin=0cm,innertopmargin=.6cm,innerbottommargin=0.2cm]
\includegraphics[width=.85\textwidth]{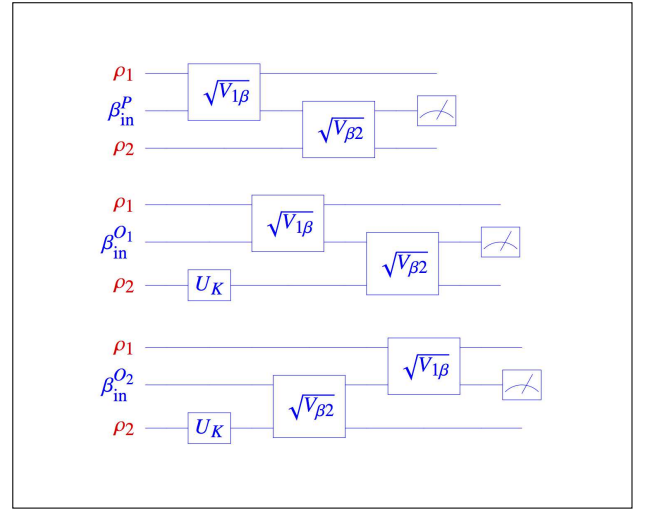}
 \end{mdframed}
  \caption[ExpScheme]{(Colors online) Alternative scheme for the detection of quantum coherence (full details in the proof). Here $d$ projective measurements are performed on an ancillary qudit. The state is not directly accessible (depicted in red), while the network elements (blue) are built at our convenience. {\bf TOP:} The network evaluates the state purity $\text{Tr}[\rho^2]$. A $\sqrt{V_{1\beta}}$ gate is applied to an ancilla  $\beta^P_\text{in}$ and a copy of the state $\rho_1$, followed by a second  $\sqrt{V_{\beta2}}$ gate applied to the ancilla and the second state copy. Projective measurements on an arbitrary basis $\ket{i}\bra{i}, i=1,\dots, d$ in the output state of the ancilla $\beta^P_\text{out}$, and on the input state (not depicted),  estimate the mean value of the purity $\text{Tr}[\rho^2]$. {\bf  CENTRE and BOTTOM:} The same scheme is employed but a copy of the state is rotated by the unitary gate $U_K$ before the interaction with the ancilla. The scheme is repeated by switching the two target states. Projective measurements at the output state of the ancilla $\beta^{O_1,O_2}_{\text{out}}$, on the initial state and on the rotated state (not depicted), determine the mean value of the overlap  $\text{Tr}[\rho U_K\rho {U_K^{\dagger}}]$.}
\label{scheme2}
\end{figure}

{\it Experimental proposals --}  In laboratory, functionals of the state density matrix are estimated by implementing programmable measurements on an ancillary qubit \cite{nielsen2,brun,paz,dariano,ekertdir,filip}. The method has been applied to measure entanglement and general quantum correlations without state reconstruction \cite{mintert,ourexp}. Here I employ it to evaluate the quantum coherence of a state whose density matrix is unknown.\\
 The square root terms prevent from recasting the skew information as a function of observables. Nevertheless, it is possible to set a non-trivial lower bound. One has $1/2 \text{Tr}\Big[[\rho,K]^2\Big]\geq \text{Tr}\Big[[\sqrt{\rho},K]^2\Big] , \forall \rho, K$, and therefore 
  \begin{align}\label{lowerbound}
{\cal I}(\rho,K)&\geq {\cal I}^L(\rho,K) \geq 0,\nonumber\\
{\cal I}^L(\rho,K)&=-1/4\ \text{Tr}\Big[[\rho,K]^2\Big].
\end{align}
Given the spectral decomposition $\rho= \sum_i \lambda_i \ket{\psi_i}\bra{\psi_i}$, the two quantities read ${\cal I}(\rho, K)= 1/2\sum_{ij}(\sqrt{\lambda_i} -\sqrt{\lambda_j})^2 K_{ij}^2, {\cal I}^L(\rho, K)= 1/4\sum_{ij}(\lambda_i -\lambda_j)^2 K_{ij}^2, K_{ij}=|\bra{\psi_i}K|\psi_j\rangle|$. The inequality is satisfied if $(\sqrt{\lambda_i}-\sqrt{\lambda_j})^2\geq 1/2 (\lambda_i-\lambda_j)^2, \forall i,j $. Simplifying, one obtains $\sqrt{\lambda_i}+\sqrt{\lambda_j} \leq \sqrt{2}$, which is always true. Also,  ${\cal I}^L(\rho,K)=0\Leftrightarrow {\cal I}(\rho,K)=0$. Note that for pure states  ${\cal V}(\rho,K)={\cal I}(\rho,K)=2{\cal I}^L(\rho,K)$, while for two-dimensional systems (qubits) the inequality $2{\cal I}^L(\rho,K)\geq {\cal I}(\rho,K)$ holds. \\
The lower bound is experimentally measurable. By defining the unitary transformation $U_K(t)=e^{i K t}$ and calculating the Taylor expansion about $t=0$, one has $\text{Tr}[\rho U_K(t) \rho U_K^{\dagger}(t)]=\text{Tr}[\rho^2]- (\text{Tr}[\rho^2 K^2]-\text{Tr}[\rho K\rho K])t^2 + O(t^3)$, and then ${\cal I}^L(\rho,K)=\frac{1}{2 t^2}(\text{Tr}[\rho^2] -\text{Tr}[\rho U_K(t) \rho U_K^{\dagger}(t)])+ O(t)$.  The two terms admit an expression in terms of observables. The purity equals the mean value of the SWAP operator $V_{AB}=\sum_{ij}\ket{i_A j_B}\bra{j_A i_B}$ applied to two state copies $\rho_{1,2}\equiv \rho$: $\text{Tr}[\rho^2]=\text{Tr}\Big[V_{12}(\rho_1\otimes\rho_2)\Big]$ \cite{brun,paz,dariano,ekertdir,filip}. On the same hand, the overlap is given by  $\text{Tr}\Big[\rho U_K \rho  {U_K^{\dagger}}\Big]=\text{Tr}\Big[V_{12}\Big(\rho_1\otimes U_{K,2}\rho_2{U_{K,2}^{\dagger}}\Big)\Big]$. 
\begin{figure} 
 \begin{mdframed}[leftmargin=0.2cm,rightmargin=0.2cm,
innerleftmargin=.5cm,innerrightmargin=0cm,innertopmargin=0.2cm,innerbottommargin=0.2cm]
\includegraphics[width=.95\textwidth]{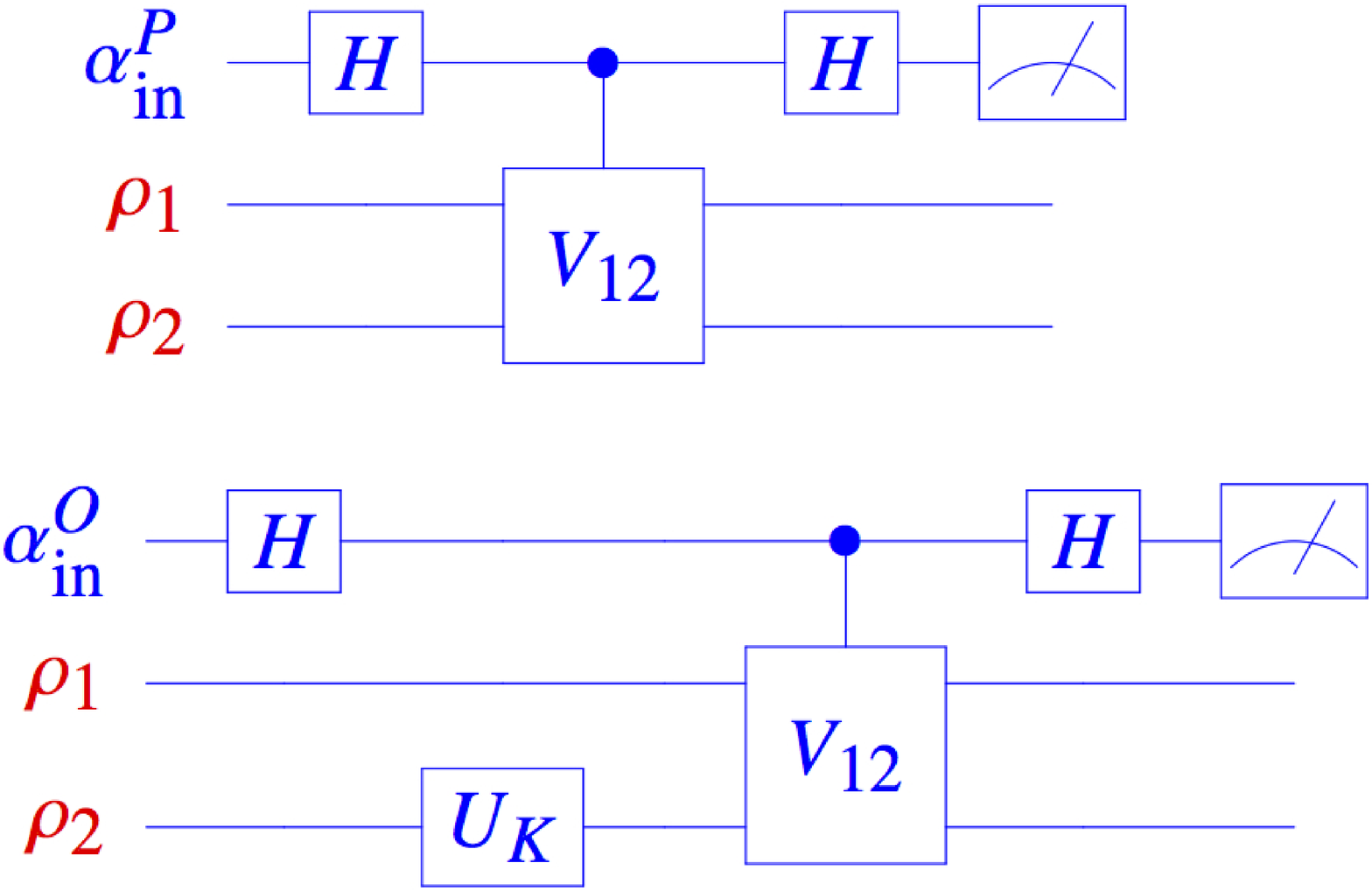}
\end{mdframed}
\caption[ExpScheme]{(Colors online) Detection of quantum coherence. The experiment consists in performing two programmable measurements on an ancillary qubit in an interferometric configuration. The density matrix of the state is not directly accessible (depicted in red), while the other elements (blue) are built at our convenience. {\bf TOP:} The network evaluates the state purity $\text{Tr}[\rho^2]$.  An ancillary control qubit  in the initial state $\alpha^{P}_{\text{in}}$  undergoes the application of a Hadamard gate  $H=\frac{1}{\sqrt{2}}\left(\begin{array}{cccc}1&1\\1&-1\end{array}\right)$, followed by an interacting  controlled-$V_{12}$ gate applied to the ancilla and the state copies: $C_{V_{12}}= \left(\begin{array}{c|c}\mathbb{I}_{d^2} & 0_{d^2} \\ \hline 0_{d^2} &   V_{12}\end{array}\right), V_{12}=\frac{1}{d}(\mathbb{I}_{d^2}+\frac{1}{d-1}\sum_i\tau_i\otimes\tau_i)$, where $\{\tau_i\}$ are the normalized $d$-dimensional Gell-Mann matrices $\{\tilde{\sigma}_i\}$: $\tau_i=\sqrt{\frac{d (d-1)}{2}}\tilde{\sigma}_i$. The SWAP can be recast in terms of projectors $P_{12}^{\pm}=\frac{1}{2}(\mathbb{I}_{d^2}\pm V_{12})=\frac{d\pm1}{2d}\mathbb{I}_{d^2}\pm \frac{1}{2(d-1)}\sum_i \tau_i\otimes\tau_i$ on the (anti)-symmetric subspaces, which are employable observables in optical setups.  Note also that any $d$-gate is decomposable in a sequence of one-qubit and two-qubit controlled-NOT transformations \cite{divincenzo}.  A second Hadamard gate is finally applied to the ancilla.  
The mean value of the ancilla polarisation, which corresponds to the visibility of the interferometer, is given by $\langle\sigma_z\rangle_{\alpha^P_{\text{out}}}= \text{Tr}[\alpha^P_{\text{in}}\sigma_z]\text{Tr}[V_{12}\rho_1\otimes\rho_2]=\text{Tr}[\alpha^P_{\text{in}}\sigma_z] \text{Tr}[\rho^2]$. {\bf BOTTOM:} The very same scheme is applied but  a copy of the state is rotated by the unitary gate $U_K$ before the interaction is switched on. 
 The ancilla polarisation is then  $\langle\sigma_z\rangle_{\alpha^O_{\text{out}}}= \text{Tr}[\alpha^O_{\text{in}}\sigma_z]\text{Tr}[V_{12}(\rho_1\otimes U_{2,K}\rho_2 U^{\dagger}_{2,K})]= \text{Tr}[\alpha^O_{\text{in}}\sigma_z] \text{Tr}[\rho_1U_{2,K}\rho_2 {U^{\dagger}_{2,K}}]$. }
\label{scheme}
\end{figure}
  The mean value of the SWAP is estimated by  implementing the interferometers in Fig.~\ref{scheme}, where an ancillary qubit  prepared in the arbitrary states $\alpha^{P,O}_{\text{in}}$ acts as the control state. Adding a controlled-SWAP gate, the polarisation of ancilla at the output  gives the mean value of the SWAP: $ \langle\sigma_z\rangle_{\alpha^P_{\text{out}}}= \text{Tr}[\alpha^P_{\text{in}}\sigma_z] \text{Tr}\Big[V_{12}(\rho_1\otimes\rho_2) \Big],  \langle\sigma_z\rangle_{\alpha^O_{\text{out}}}= \text{Tr}[\alpha^O_{\text{in}}\sigma_z] \text{Tr}\Big[V_{12}\Big(\rho_1\otimes U_{2,K}\rho_2 {U_{2,K}}^{\dagger}\Big) \Big]$. Hence,
   \begin{Res}\label{bound}
 The experimental evaluation of (a lower bound of the) quantum coherence of an unknown state in a $d$-dimensional system requires two programmable measurements on an ancillary qubit.
\end{Res}
Quantum coherence is measurable by means of two measurements only, while tomographic state reconstruction would require $O(d^2)$ direct measurements on the system. For quantum gates acting on qubits, i.e. the building blocks of quantum algorithms, any observable is defined by $K=\vec{n}\cdot\vec{\sigma}, |\vec{n}|=1$, being  $\vec{\sigma}=\{\sigma_i\}$ the Pauli matrices. One thus obtains the simplified expression ${\cal I}^L(\rho,K)=\frac{1}{2}(\text{Tr}[\rho^2]-\text{Tr}[\rho U_K(\theta)|_{\theta=\pi/2} \rho U_K^{\dagger}(\theta)|_{\theta=\pi/2}])$. \\
 The controlled gate may be cumbersome to implement. It is then useful to work out an alternative scheme. It is known that the purity can be evaluated by applying twice the  ``$\sqrt{\text{SWAP}}$''  operator $\sqrt{V_{AB}}=\frac{1}{\sqrt2}(\mathbb{I}_{d^2} -  i\  V_{AB})$ in parallel to the ancilla and each copy of the state  \cite{pasca}. I generalize such a protocol to measure the overlap of two arbitrary states, and to build an alternative detection scheme of quantum coherence (Fig. \ref{scheme2}, proof at the end of the main text). The outcomes of projective measurements over a basis $\{\ket{i}\bra{i}\}, i=1,\ldots,d,$ made on the output state of the ancilla in each of the three interferometric configurations in Fig.~\ref{scheme2}, with additional measurements on the state and the rotated state, $\text{Tr}[X\ket{i}\bra{i}], X=\beta^{P,O_1,O_2}_{\text{out}}, \rho, U_K \rho U_K^{\dagger}, i=1,2,\ldots, d  $,  determine both the purity and overlap terms. In conclusion:
 \setcounter{Res}{1}\label{bound2}
\renewcommand{\theRes}{\arabic{Res}/bis}
 \begin{Res}
 The experimental detection of (a lower bound of the) quantum coherence of an unknown state in a $d$-dimensional system requires $O(d)$ projective measurements on an ancillary qudit and the system itself.
\end{Res}
  The strategy still enjoys a polynomial advantage against state tomography.\\

{\it Discussion --} 
 I introduced a model-independent quantitative characterization of quantum coherence for states of finite dimensional systems. At theoretical level, the skew information quantifies coherence as the genuinely quantum uncertainty of a measurement. The significance of the proposed experimental  schemes rests on their scalability, outperforming protocols based on state reconstruction, and generality. 
 The result also suggests a new approach, based on information geometry \cite{amari}, to study open quantum systems. I proved here that the skew information is the geometric entity which describes coherence. This quantity belongs to the family of Riemnannian metrics on the statistical manifold of quantum states which are monotonically decreasing under quantum channels \cite{petz}. Then the evolution of such a metric (and related higher order tensors) may help monitor quantum back-flow of information in non-Markovian dynamics and the supra-classical efficiency of energy transport mechanisms in biological complexes \cite{flow,biorev,nonmarkrev}, shaping our knowledge of quantum memory effects in open systems.\\
  
  \noindent{\it Proof of Result 1.}\\
a) The skew information is a faithful measure of coherence. It is convex, non-negative 
\cite{wy}, and vanishes if the only if the state is incoherent. The latter is defined as a state whose density matrix is diagonal in a given basis. 
By definition, ${\cal I}(\rho, K)=0\Leftrightarrow[\rho,K]=0$, i.e. state and observable diagonalize in the same eigenbasis q.e.d.   \\
  b)It is monotonically non-increasing under incoherent operations, which are expressed by a set of Kraus operators $\{K_n\}$ such that $\sum_n K_n^{\dagger}K_n=\mathbb{I}, K_n^{\dagger} {\cal I}_K K_n \subset {\cal I}_K, \forall n$, where ${\cal I}_K$ is the set of incoherent states w.r.t. $\{\ket{k_i}\}$.  \\
     The skew information ${\cal I}(\rho,K)$ does not increase on average by a von Neumann measurement of $K$: $\sum_n p_n{\cal I}(K_n\rho K_n^{\dagger},K)\leq {\cal I}(\rho,K)$ \cite{frenkel}. The result and the convexity of the skew information proves the monotonicity for completely positive trace preserving (CPTP) incoherent maps, and that for any incoherent state $\rho_{{\cal I}_K}$ one has ${\cal I}(K_n^{\dagger}\rho_{{\cal I}_K}K_n,K)=0, \forall n$. I provide an alternative constructive argument for the class of $K$-invariant operations, which is a subset of the CPTP ones \cite{plenio}.  The skew information of a bipartite state $\rho_{AB}$ satisfies ${\cal I}(\rho_{AB}, K_A\otimes \mathbb{I}_B)\geq {\cal I}(\text{Tr}_B[\rho_{AB}], K_A), \forall K$. A $K$-invariant channel on a system $A$ takes the form ${\cal E}^{K}_A(\rho_A)=\text{Tr}_B[V^{K}_{AB}(\rho_A\otimes\tau_B)V^{K\dagger}_{AB}]$, where $V^K_{AB}$ is a $K$-invariant unitary, i.e. $V^K_{AB} (K_A\otimes\mathbb{I}_B+\mathbb{I}_A\otimes K_B)V^{K\dagger}_{AB}=K_A\otimes\mathbb{I}_B+\mathbb{I}_A\otimes K_B$, and $\tau_B\in {\cal I}_K$. One then obtains ${\cal I}(\rho_A, K_A)={\cal I}(\rho_A\otimes\tau_B, K_A\otimes\mathbb{I}_B+\mathbb{I}_A\otimes K_B)={\cal I}(\rho_A\otimes\tau_B, V^K_{AB}(K_A\otimes\mathbb{I}_B+\mathbb{I}_A\otimes K_B)V^{K\dagger}_{AB})={\cal I}(V^{K\dagger}_{AB}(\rho_A\otimes\tau_B)V^{K}_{AB}, K_A\otimes\mathbb{I}_B+\mathbb{I}_A\otimes K_B)\geq {\cal I}(\text{Tr}_B[V^{K\dagger}_{AB}(\rho_A\otimes\tau_B)V^{K}_{AB}], K_A)={\cal I}({\cal E}_A^{K_A}(\rho_A),K_A)$ q.e.d. \\
  One may further demand monotonicity under classical encoding: $ {\cal I}(\sum_n p_n K_{n,A}^{\dagger}\rho_A K_{n,A}\otimes \ket{n}\bra{n}_B, K_A\otimes \mathbb{I}_B)\leq {\cal I}(\rho_A,K_A), \ket{n}\bra{n}\in {\cal I}_K$ (criterion C2c of \cite{plenio}). The property is satisfied, since ${\cal I}(\sum_n p_n K_{n,A}^{\dagger}\rho_A K_{n,A}\otimes \ket{n}\bra{n}_B, K_A\otimes \mathbb{I}_B) \leq \sum_n p_n{\cal I}(K_{n,A}^{\dagger}\rho_A K_{n,A}, K_A) \leq {\cal I}(\rho_A, K_A)$ q.e.d. \\

\noindent{\it Proof of Result 2/bis.}\\
Here I prove that the schemes in  Fig. \ref{scheme2} evaluate the overlap of two arbitrary density matrices $\rho_{A,B}$, generalizing  Ref.~\cite{pasca}. Result 2/bis is then a case study with $\rho_{A,B}=\rho$ (TOP scheme), $\rho_A=\rho, \rho_B=U_K \rho U_K^{\dagger}$ (CENTRE) and $\rho_A=U_K \rho U_K^{\dagger}, \rho_B=\rho$ (BOTTOM).   The steps of the protocol are: \\
a) Preparation of the input states: a $d$-dimensional ancilla (in a pure state, for simplicity) and two $d$-dimensional states whose density matrices are respectively $\beta=\frac{1}{d}(\mathbb{I}_d+\vec{x}_{\beta}\cdot \vec{\tau}), |\vec{x}_{\beta}|=1,\rho_A=\frac{1}{d}(\mathbb{I}_d+\vec{x}_{A}\cdot \vec{\tau}), \rho_B=\frac{1}{d}(\mathbb{I}_d+\vec{x}_{B}\cdot \vec{\tau})$. The goal is to determine $\text{Tr}[\rho_A \rho_B]=\frac1d(1+ (d-1)\vec{x}_A\cdot\vec{x}_B)$.\\
b)Application of the gate  $\sqrt{V_{A\beta}}=\frac{1}{\sqrt2}(\mathbb{I}_{d^2} -   i V_{A\beta})$ to the state $\rho_A$ and the ancilla $\beta$. The resulting marginal state of the ancilla at this intermediate stage is given by $\beta_{\text{int}}=\frac{1}{d}(\mathbb{I}_d+ \vec{y}_{\beta} \cdot \vec{\tau}),$
where $\vec{y}_{\beta}=\frac{1}{2}\Big(\vec{x}_{A}+\vec{x}_{\beta}+(d-1)\vec{x}_{A}\wedge \vec{x}_{\beta}\Big)$ and $\wedge$ is the exterior product.\\
c)Implementation of the second $\sqrt{V_{\beta B}}$ gate to the ancilla and the state $\rho_B$. The output state of the ancilla reads $\beta_{\text{out}}=\frac{1}{d}(\mathbb{I}_d+ \vec{z}_{\beta} \cdot \vec{\tau})$, with $\vec{z}_{\beta}=\frac{1}{2}\Big(\vec{x}_{B}+\vec{y}_{\beta}+ (d-1) \vec{y}_{\beta}\wedge \vec{x}_{B}\Big)$.\\
d)Performing a complete set of  $d$ projective measurements over a basis $\{\ket{i}\bra{i}=\rho_i, i=1,2,\ldots d$ on the output state of the ancilla. A clever choice is such that the pure state $\beta=\ket{i_\beta} \bra{i_{\beta}}$ is an element of the basis: $\text{Tr}[\beta \ket i\bra i]=\delta_{i i_{\beta}}$. The outcome of each measurement is $S^i_{AB}=\text{Tr}[\beta_{\text{out}}\vec{x}_i\cdot\vec{\tau}]=(d-1)\vec{z}_{\beta}\cdot\vec{x}_i= \frac{d-1}{2}\Big(\vec{x}_B\cdot \vec{x}_i+\frac{\vec{x}_A\cdot\vec{x}_i+\vec{x}_{\beta}\cdot\vec{x}_i}{2}\Big)+ \frac{(d-1)^2}{4}(\vec{x}_A\wedge\vec{x}_{\beta}\cdot \vec{x}_i+\vec{x}_A\wedge\vec{x}_{B}\cdot \vec{x}_i+\vec{x}_{\beta}\wedge\vec{x}_{B}\cdot \vec{x}_i)+\frac{(d-1)^3}{4}((\vec{x}_A\wedge\vec{x}_{\beta})\wedge\vec{x}_{B}\cdot\vec{x}_i)$.\\
e)Repetition of  the protocol by interchanging $\rho_A,\rho_B$, obtaining the term $S^i_{BA}$.  One then has $S^{i}_{AB}+S^i_{BA}=\frac{d-1}{4}(3 (\vec{x}_A\cdot\vec{x_i}+ \vec{x}_B\cdot\vec{x_i})+2 \delta_{i i_{\beta}})+\frac{(d-1)^3}{4}((\vec{x}_A\wedge\vec{x}_{\beta})\wedge\vec{x}_{B}\cdot\vec{x}_i +(\vec{x}_B\wedge\vec{x}_{\beta})\wedge\vec{x}_{A}\cdot\vec{x}_i).$ After some algebra (see appendix of \cite{pasca} for the case $A=B$), one obtains $(\vec{x}_A\wedge\vec{x}_{\beta})\wedge\vec{x}_{B}\cdot\vec{x}_i +(\vec{x}_B\wedge\vec{x}_{\beta})\wedge\vec{x}_{A}\cdot\vec{x}_i=\frac{1}{(d-1)^2}(2 (\vec{x}_A\cdot \vec{x}_B) \delta_{i i_{\beta}}-(\vec{x}_A\cdot \vec{x}_i)(\vec{x}_B\cdot \vec{x}_{i_{\beta}})-(\vec{x}_B\cdot \vec{x}_i)(\vec{x}_A\cdot \vec{x}_{i_{\beta}}))$ \\
f)Additional $d$ projective measurements on $\rho_{A,B}$ have outcomes $S^i_{A,B,\beta}=(d-1)\vec{x}_{A,B,\beta}\cdot \vec{x_i}$. 
  The overlap is then determined by:
\begin{eqnarray}
\vec{x}_A\cdot\vec{x}_B&=&\sum_{i=1}^d \Big(2(S^i_{AB}+S^i_{BA})-3/2 (S^i_A+S^i_B) \nonumber\\
&+&\frac{1}{2(d-1)}(S^i_AS^{i_{\beta}}_B+S^i_B S^{i_{\beta}}_A)\Big)-1. 
\end{eqnarray}
The method requires $5 d$ measurements (to obtain $S_{AB},S_{BA},S_{A},S_B$, for the overlap and $S_{AA}$ for the purity). Allowing for interacting gates between $\rho_{A,B}$, the task requires $4 d$ measurements. In such a case, the protocol has to be run setting $\beta=\rho_B=\rho, \rho_A=U_K\rho {U_K}^{\dagger}$, then switching to $\beta=\rho_B=U_K\rho {U_K}^{\dagger}, \rho_A=\rho$ and finally making $d$ projective measurements on $\rho_{A,B}$. 



\section*{Acknowledgments}
This work was supported by the Singapore National Research
Foundation under NRF Grant No. NRF-NRFF2011-07 and the EPSRC grant EP/L01405X/1.
I thank  B. Aaronson, G. Adesso,  M. Ahmadi, D. Brody,  A. Chaudry, T. Farrow, P. Gibilisco, F. Hansen, I. Mekhov,  R. Nair, M. Tsang, T. Tufarelli and V. Vedral for fruitful discussions.

   \newpage
   
 \appendix*
\begin{widetext}

\section{Supplemental Material}
\begin{center}
{\large{\bf {Observable measure of quantum coherence in finite dimensional systems}}}

\quad \\

{\normalsize Davide Girolami}

\end{center}

\setcounter{equation}{0}

\section{A. Quantum coherence in multipartite systems}
The uncertainty on a quantum measurement is affected by the correlations shared by the system of interest with other parties. While absence of entanglement does not entail classicality, as it is not necessary to ensure coherence, the concepts of quantum coherence and quantum discord \cite{OZ,HV}, are  entwined.  The latter is defined as the least amount of disturbance experienced by a compound system due to a local measurement on one of the subsystems, say $A$.  If and only if discord-like correlations are shared among parts of a compound system, then quantum coherence is guaranteed in {\it any} local basis \cite{genio}. Indeed, a class of {\it bona fide} measures of quantum discord of state $\rho_{AB}$ is defined by $\min\limits_{K^{\Gamma}_A}{\cal I}(\rho_{AB},K_A^{\Gamma}\otimes\mathbb{I}_B)$, where $\Gamma$ indexes the spectrum of  $K_A^{\Gamma}$ \cite{tufas}. The minimization is made over same spectrum observables, and each choice of the spectrum pinpoints a specific measure.\\
Furthermore, the skew information framework highlights a to date unexplored (to my knowledge) kind of statistical dependence. Here I define the {\it residual $K_{A_i}$-coherence} as the difference between the quantum coherence of global and marginal states:
\begin{eqnarray}
{\cal C}(\rho_{A_1,A_2,\ldots, A_n},\mathbb{I}_{A_i,A_2,\ldots, A_{i-1}}\otimes K_{A_i}\otimes \mathbb{I}_{A_{i+1},A_{i+2},\ldots,A_n})={\cal I}(\rho_{A_1,A_2,\ldots, A_n},\mathbb{I}_{A_i,A_2,\ldots, A_{i-1}}\otimes K_{A_i}\otimes \mathbb{I}_{A_{i+1},A_{i+2},\ldots,A_n})-{\cal I}(\rho_{A_i},K_{A_i}),
\end{eqnarray}
where $\rho_{A_i}$ is the marginal state of subsystem $A_i$. The quantity is nonnegative, vanishing for states whose density matrix is block-diagonal in the eigenbasis of $\mathbb{I}_{A_i,A_2,\ldots, A_{i-1}}\otimes K_{A_i}\otimes \mathbb{I}_{A_{i+1},A_{i+2},\ldots,A_n}$.
The operational power and a full algorithmic characterization of the residual $K_{A_i}$-coherence is worthy of investigation.
 
\section{B. The skew information is a measure of asymmetry}

 The absence of a reference frame has been proven equivalent to constrain quantum dynamics by a superselection rule (SSR) \cite{bartsup,review,www,ahr,ahr2,wightman,new}, while  the ability of a system to act as reference frame is the quantum resource known as  asymmetry or frameness \cite{review}. Here I provide the proof that the skew information is a measure of asymmetry. For the sake of clarity, I recall the technical definition of  SSR \cite{www,ahr,ahr2,bartsup}.\\
  A $G$-SSR for a quantity $Q$ (supercharge) is defined as a law of invariance of the state of a system with respect to a transformation group $G$. Given a system with Hilbert space ${\cal H}$ and a unitary representation $U: G \rightarrow{\cal B}(\cal H)$ mapping the group to the set of bounded observables on the Hilbert space, any operation ${\cal E}^G$ is said $G$-covariant if it satisfies ${\cal E}^G (U(g)\rho U(g)^{\dagger})= U(g){\cal E}^G(\rho ) U(g)^{\dagger}, \forall g \in G$.  There is no way to distinguish by means of a $G$-covariant operation, without violating the $G$-SSR, the state $\rho$ from $U(g)\rho U(g)^{\dagger}, \forall g$. Thus, for finite groups, the physical states are described by the density matrices obtained by averaging over the group transformations through the $G$-twirling operation  ${\cal G}[\rho]=\frac{1}{\text{dim}\ G}\sum_{g\in G} U(g)\rho U(g)^{\dagger}$ (an equivalent definition holds for Lie groups) \cite{bartsup}. Any density matrix $\rho$ with off-diagonal entries (coherence) in the basis of the eigenstates of $Q$ is projected by the average over the group transformations into the diagonal state ${\cal G}[\rho]$.  Consequently, it is not distinguishable, by allowed physical operations, from  ${\cal G}[\rho]$, and it cannot be exploited for quantum information tasks, unless one could overcome the limitations imposed by the SSR by accessing a reference frame, i.e. an ancillary system $R$ which shows coherent superpositions of supercharge eigenstates. \cite{bartsup,ahr,ahr2,review}.  The only states left invariant by the $G$-twirling are either eigenstates $|q\rangle$ of $Q$ or mixtures of its eigenstates $\sum_q c_q |q\rangle\langle q|, \sum_q c_q=1$.  \\
An entropic measure of asymmetry is the relative entropy of $G$-frameness or $G$-asymmetry ${\cal S}({\cal G}[\rho])-{\cal S}(\rho)$, being ${\cal S}$ the von Neumann entropy \cite{goursup,gour2sup,vacca2sup} . The skew information is another consistent measure of quantum asymmetry \cite{marvianthesis}:
\begin{Rem}\label{symres}
Given a $G$-SSR with supercharge $Q$, the skew information ${\cal I}(\rho, Q)= -\frac12$\normalfont{Tr}$\Big[[\sqrt{\rho},Q]^2\Big]=$ \normalfont{Tr}$[\rho Q^2 - \sqrt{\rho}Q\sqrt{\rho}Q]$ satisfies the criteria identifying an asymmetry measure of the state $\rho$ \cite{marvianthesis,goursup}. 
\end{Rem}
\begin{proof}a) The skew information is a faithful measure of asymmetry. It is convex, non-negative  and ${\cal I}(\rho, Q)=0 \Leftrightarrow \rho = {\cal G}[\rho]$. Under a SSR, a physical state $\rho$ is either eigenstate $|q\rangle$ of $Q$ or mixture of its eigenstates $\sum_q c_q |q\rangle\langle q| $. In the first case, it is trivial to see that the skew information is zero, while for the mixture, by exploiting the convexity of ${\cal I}$, one obtains ${\cal I}(\sum_q c_q |q\rangle\langle q|,Q)\leq \sum_q c_q {\cal I}(|q\rangle\langle q|,Q)=0$. Also, by construction one has ${\cal I}(\rho, Q)=0  \Leftrightarrow [\rho,Q]=0$ and $  [\rho,Q]=0 \Leftrightarrow [\rho, U(g)]=0, \forall g \in G$, which is true if and only if the state is symmetric, i.e.  $\rho={\cal G}[\rho]$ \cite{review}. \\
b) It is monotonically non-increasing under $G$-covariant operations: ${\cal I}({\cal E}^G(\rho), Q)\leq {\cal I}(\rho, Q)=0, \forall {\cal E}^G$. $G$-covariant operations correspond to a subset of  incoherent operations with respect to the basis $\{|q\rangle\}$. The proof b) of Res. 1 works here as well. In the very same way, one builds $G$-covariant operations ${\cal E}^{{\cal I}_Q}_A$ and shows that ${\cal I}(\rho_A,Q_A)\geq {\cal I}({\cal E}_A^{{\cal I}_Q}(\rho_A),Q_A)$ (see Theorem II.1 of \cite{rudosup}).\\ 
 \end{proof}
 
Quantum asymmetry represents the amount of coherence in the eigenbasis of the supercharge \cite{new,goursup}. Under a SSR on the system $S$, one has ${\cal I}(\rho_{S}, Q_S)= 0$. It is known that the SSR is broken by introducing a reference frame $R$, which has access to asymmetric states, and then coupling it by a global $G$-invariant operation with $S$ \cite{review}.  I give a general quantitative prescription for  symmetry breaking transformations in terms of quantum uncertainties, by exploiting the properties of the skew information. For any observables $Q_S, Q_R$ and states $\rho_S, \tau_R$, one has ${\cal I}({\rho}_{S}\otimes\tau_R, Q_S\otimes\mathbb{I}_R+ \mathbb{I}_S\otimes Q_R)={\cal I}(\rho_{S}, Q_S )+ {\cal I}(\tau_R, Q_R)$ and ${\cal I}(\rho_{SR}, Q_S\otimes\mathbb{I}_R+ \mathbb{I}_S\otimes Q_R)=0 \Rightarrow {\cal I}(\text{Tr}_R[\rho_{SR}], Q_S)={\cal I}(\text{Tr}_S[\rho_{SR}], Q_R)= 0$. Given the $G$-invariant transformation $U_{SR}(\rho_S\otimes\tau_R) U_{SR}^{\dagger}=\tilde{\rho}_{SR}$, one obtains ${\cal I}(\tilde{\rho}_{SR}, Q_S\otimes\mathbb{I}_R+ \mathbb{I}_S\otimes Q_R)={\cal I}(\rho_{S}\otimes\tau_{R}, Q_S\otimes\mathbb{I}_R+ \mathbb{I}_S\otimes Q_R)={\cal I}(\tau_R,{\cal Q}_R)$.  Thus, if ${\cal I}(\tau_R, Q_R)> 0$, one can obtain  asymmetric states of the system $S$: ${\cal I}(\text{Tr}_R[\tilde{\rho}_{SR}], Q_S)>0$. \\
  I finally remark that if the observable is the Hamiltonian of the system, then the skew information turns out to be the Hessian matrix of the relative entropy between the state and the equilibrium state, which equals the free energy \cite{donald}.  The role played by quantum coherence and asymmetry in the thermodynamics of quantum systems certainly deserves to be explored.

\end{widetext}

\clearpage

\end{document}